\documentstyle[epsfig]{qcdparis_mod}
\def\a0{\bar\alpha_0}

\def\ap{\bar\alpha_{r-1}}
\def\as{\alpha_{\mbox{\tiny S}}}
\def\b0{\beta_0}
\def\BFKL{{\mbox{\tiny BFKL}}}
\def\CCFM{{\mbox{\tiny CCFM}}}

\def\DelN{\Delta_{\mbox{\tiny N}}}
\def\DelR{\Delta_{\mbox{\tiny R}}}
\def\Ecm{E_{\mbox{\scriptsize cm}}}
\def\ee{e^+e^-}

\def\HW{{\small HERWIG}}
\def\jet{{\mbox{\scriptsize jet}}}
\def\kt{k_\bot}
\def\lab{{\mbox{\scriptsize lab}}}
\def\lef{\Lambda_{\mbox{\scriptsize eff}}}
\def\lms{\Lambda^{(5)}_{\overline{\mbox{\tiny MS}}}}

\def\MSbar{\overline{\mbox{MS}}}
\def\NCS{N_{\mbox{\tiny CS}}}
\def\pt{{\mbox{\scriptsize pert}}}
\def\pw{{\mbox{\scriptsize pow}}}
\def\sds{{\mbox{\tiny DIS}}}
\def\sdy{{\mbox{\tiny DY}}}
\def\see{{\mbox{\scriptsize ee}}}
\def\sL{\sigma_{\mbox{\tiny L}}}
\def\st{\sigma_{\mbox{\scriptsize tot}}}

\def\yc{y_{\mbox{\scriptsize cut}}}
\def\frac#1#2{ {{#1} \over {#2} }}

\def\VEV#1{\left\langle #1\right\rangle}
\def\beq{\begin{equation}}
\def\beeq{\begin{eqnarray}}
\def\eeq{\end{equation}}
\def\eeeq{\end{eqnarray}}
\begin{document}
\pagestyle{plain}
\title{Hadronic Final States$\,^\ast$}

\author{B.R.\ Webber}

\affil{Cavendish Laboratory, University of Cambridge,\\
       Madingley Road, Cambridge CB3 0HE, U.K.}

\abstract{
The following aspects of hadronic final states in deep inelastic
lepton scattering are reviewed:
measuring $\as$ from multi-jet production rates and event shapes;
alternative jet algorithms for DIS;
power-suppressed corrections to event shapes;
comparing jet fragmentation in $\ee$ annihilation and DIS;
final states in the BFKL and CCFM formulations of small-$x$ dynamics;
exotic (instanton-induced) final states.
}

\resume{
Les aspects suivants des \'etats finals hadroniques etc.
}

\twocolumn[\maketitle]
\fnm{7}{Plenary talk at the Workshop on Deep Inelastic Scattering
and QCD, Paris, April 1995.  Research supported in part by the UK
Particle Physics and Astronomy Research Council and by the EC
Programme ``Human Capital and Mobility", Network ``Physics at
High Energy Colliders", contract CHRX-CT93-0357 (DG 12 COMA).}

\section{Introduction}
This talk will review a selection of topics concerning hadronic
final states in deep inelastic lepton scattering (DIS).
One of the primary aims of studies of DIS final states
is to test the predictions of QCD in more detail than is
provided by measurements of the totally inclusive structure
functions. By analogy with $\ee$ annihilation, it is natural
to begin by studying various global measures of the
jet-like properties of the final state. This is the topic of the
following section.  Amongst the quantities considered are the
multi-jet production rates, defined according to various jet
algorithms, and event shape parameter distributions.  Both
of these have to be defined in an infrared-safe way, so as
to be insensitive to the long-distance behaviour of QCD and
hence calculable in perturbation theory.  Recently, it has
been found that observables satisfying these conditions can
still differ substantially in their sensitivity to
non-perturbative physics, through the magnitude and energy
dependence of corrections that are suppressed by inverse
powers of a large momentum scale. Some recent theoretical
ideas on this question and their predictions for DIS will
be discussed.  Experimental and theoretical results on the
fragmentation spectra of DIS current jets will also be
reviewed.

Section 3 deals with final state features of DIS in the
region of very small Bjorken $x$ values currently being
investigated at HERA.  Here the theoretical analysis is
more difficult and a number of basic points remain to be
clarified. Some of the most promising indicators of
possible new dynamics at small $x$ concern final state
properties, such as the transverse energy flow and the
production of forward jets with transverse momenta
comparable to the DIS momentum transfer $Q$.

In section 4, a more speculative but exciting possibility
for DIS final states is discussed, namely non-perturbative
processes that might be induced by QCD instantons.  The
associated final states would have high hadron multiplicities
without any prominent multi-jet production.

Section 5 contains some concluding remarks.

\section{Jet Physics}
\subsection{Jet rates}
Multi-jet production rates are at present the only
features of DIS final states that have been fully predicted
to next-to-leading order in QCD perturbation theory
\cite{nlo}. Denoting by $\sigma_{n+1}(x,Q^2;\yc)$ the cross
section for the production of $n$ jets plus the remnant jet,
at given values of Bjorken $x$, momentum transfer-squared $Q^2$
and jet resolution $\yc$, we have schematically
\beeq
\sigma_{n+1}(x,Q^2;\yc)&=&\as^{n-1}(Q) A_n(x,Q^2;\yc)\nonumber \\
                       &+&\as^n(Q) B_n(x,Q^2;\yc)+\ldots
\eeeq
where $A_n$ and $B_n$ are calculable in terms of the parton
distribution functions of the proton.  The leading terms $A_n$
are given by the corresponding tree graphs as illustrated
in \fref{feyn}.  Once the next-to-leading term $B_n$ is known
for $n>1$, this provides a method for measuring the strong
coupling $\as$.

\ffig{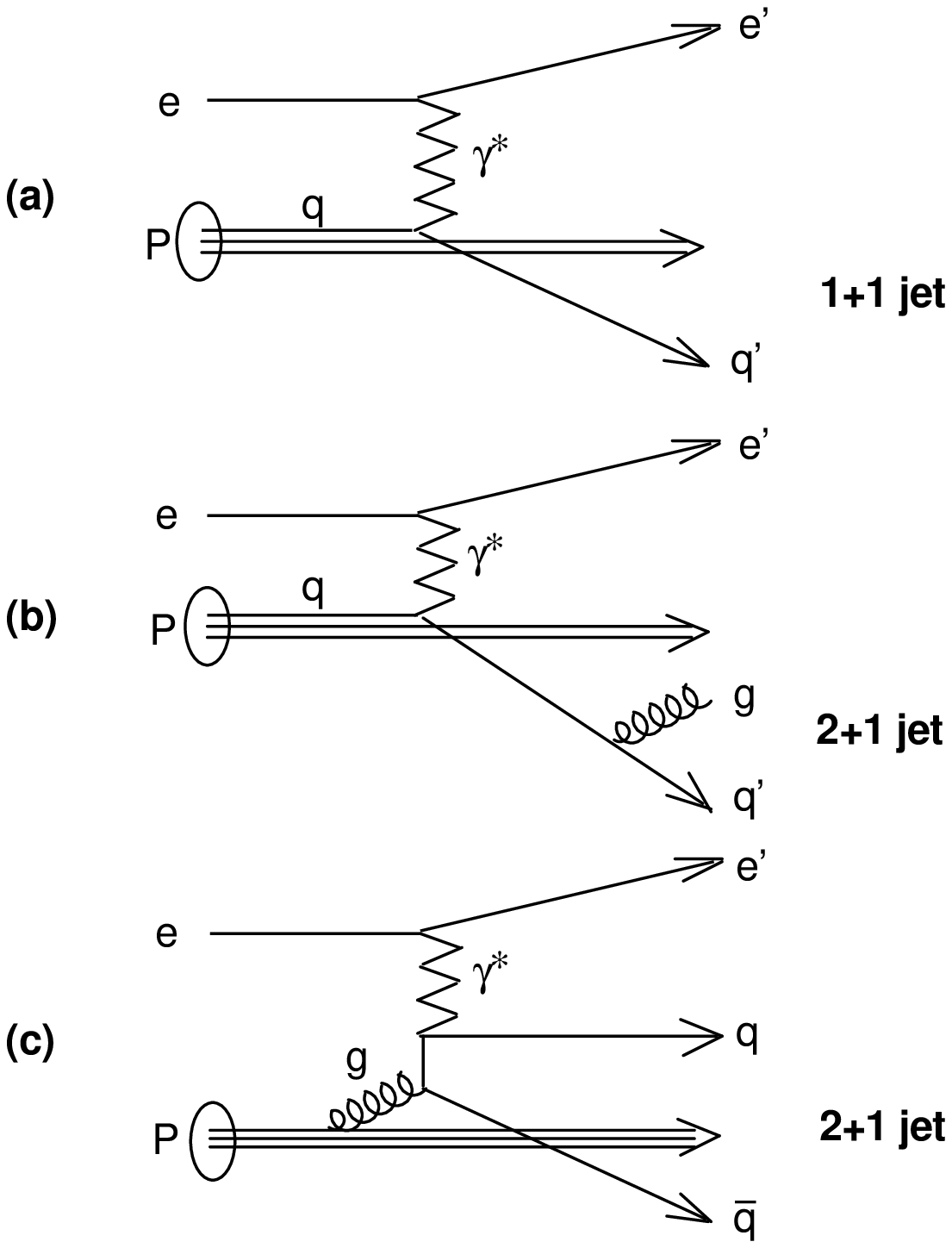}{80mm}{\em Typical diagrams for neutral
current deep-inelastic scattering: (a) Born term, (b) QCD~Compton
scattering, (c) boson-gluon fusion.}{feyn}

There are several different ways of defining jets, with correspondingly
different definitions of the jet resolution $\yc$. So far, the
next-to-leading function $B_2$ for 2+1 jets has been computed only
for the {\em modified JADE} jet clustering algorithm~\cite{nlo,jade}.
Here one defines for each pair of particles or clusters $i$ and
$j$ the quantity
\beq\label{mJjets}
y_{ij} = 2E_iE_j(1-\cos\theta_{ij})/W^2\simeq m^2_{ij}/W^2
\eeq
where $m_{ij}$ is the invariant mass of the pair and $W$ is the
overall hadronic centre-of-mass energy ($W^2=Q^2(1-x)/x$). The
pair with the smallest value of $y_{ij}$ are clustered, the
process being repeated until all $y_{ij}$'s are above $\yc$.
The proton remnant is included in the clustering procedure
as a `pseudoparticle' carrying any missing longitudinal
momentum that is lost down the beam pipe.  After clustering,
the clusters remaining at resolution $\yc$ are defined as jets,
the one containing the pseudoparticle being the remnant jet.

Using the above jet definition at $\yc=0.02$, the HERA experiments
have obtained evidence of the running of $\as$, and have measured
its value as~\cite{disjH1,disjZ}
\beq\label{asH}
\as(M_Z) = 0.123\pm 0.018\;\;\;\;\mbox{(H1),}
\eeq
\beq\label{asZ}
\as(M_Z) = 0.117\pm 0.005^{+0.004}_{-0.005}\pm 0.007
\;\;\mbox{(ZEUS),}
\eeq
where the three components of the error in (\ref{asZ}) are statistical,
experimental systematic and theoretical systematic, respectively.
In \fref{disjZfig}, for example, the ZEUS results on $\as$ as a
function of $Q^2$ are shown, together with the expected
behaviour for various values of the 5-flavour $\MSbar$
QCD scale $\lms$.  The inner error bars show the statistical
errors only, while the outer ones represent statistical
and systematic errors added in quadrature.

\ffig{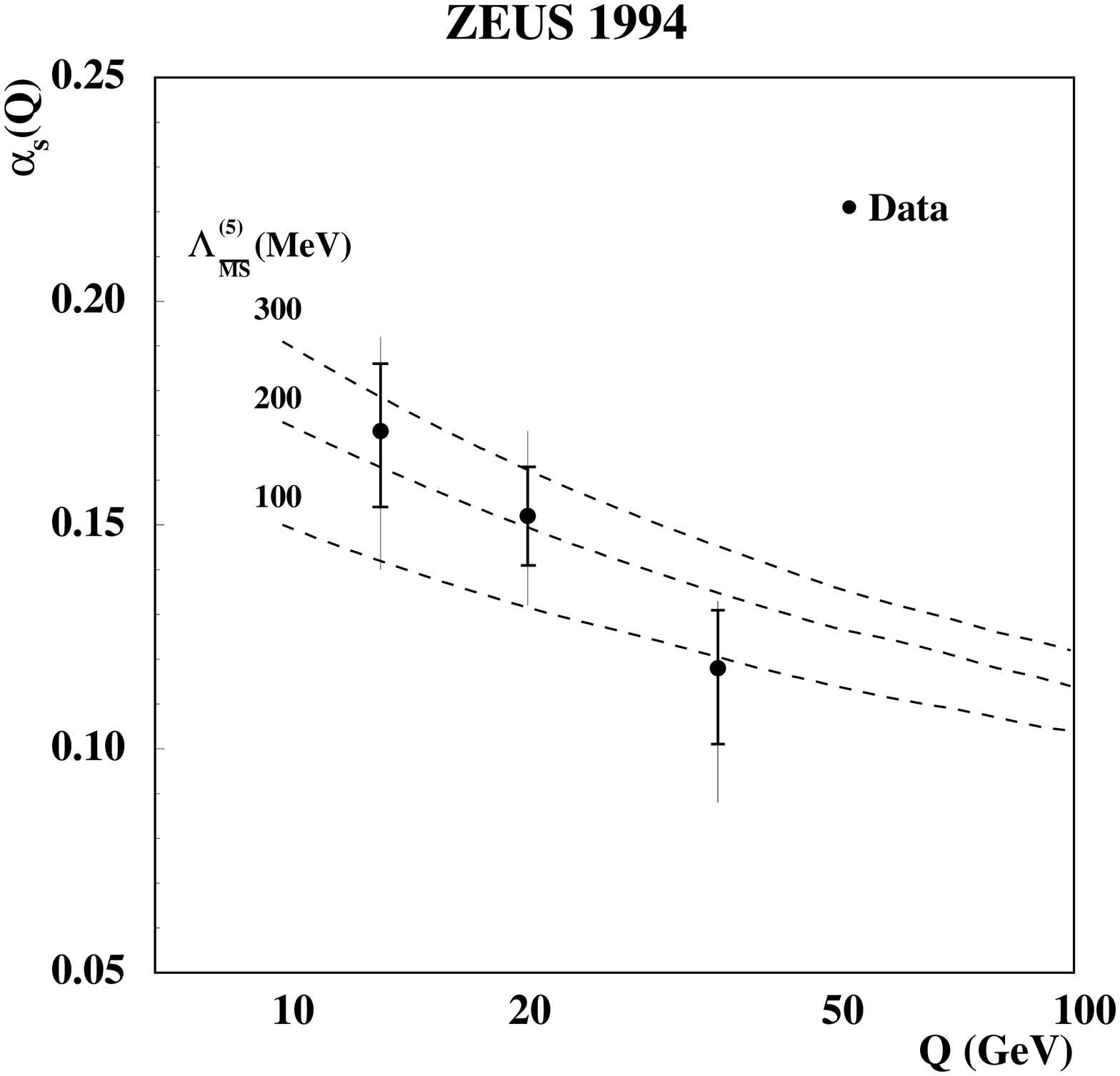}{80mm}{\em Results from the ZEUS
collaboration on $\as$ for different $Q^2$ regions.}{disjZfig}

In addition to the modified JADE algorithm used in the above
analyses, there are other possible jet definitions, which
have some theoretical advantages but are not yet predicted
to ${\cal O}(\as^2)$. For future reference I mention here
the {\em factorizable JADE} algorithm~\cite{fJADE}, with
\beq\label{fJjets}
y_{ij} = 2E_iE_j(1-\cos\theta_{ij})/Q^2\simeq m^2_{ij}/Q^2\;,
\eeq
i.e.\ differing from (\ref{mJjets}) only in the normalization,
and the $\kt$-{\em algorithm} for DIS~\cite{DISkt}, in which
\beq\label{ktjets}
y_{ij} = 2\min\{E_i^2,E_j^2\}(1-\cos\theta_{ij})/Q^2\simeq k^2_{\bot
ij}/Q^2\;,
\eeq
where $\kt$ represent the transverse momentum of $i$ relative to $j$,
or vice versa, whichever is the smaller.  In both these algorithms
the definition of the remnant jet is also different from that in
the modified JADE algorithm:  instead of the
pseudoparticle, one introduces a fixed momentum vector $p_r=xp$,
where $p$ is the incoming proton momentum, and computes $y_{ir}$,
obtained by clustering $p_i$ with $p_r$, along with $y_{ij}$ at
each stage.  If the smallest of all these is $y_{kr}$, then $k$
is classified as part of the remnant jet and is not available
for further clustering.

A substantial part of the theoretical systematic error in
the $\as$ determinations (\ref{asH}) and (\ref{asZ}) is
associated with {\em hadronization corrections},
i.e. the non-perturbative corrections applied to
the perturbative predictions before comparing them
with hadron-level data.  The $\kt$-algorithm in particular
may have some advantages in reducing this source of error.
As we shall see in section~2.3, there are theoretical
indications that non-perturbative effects should
be smaller for algorithms based on the $\kt$-resolution
variable (\ref{ktjets}) rather than the JADE resolution
(\ref{mJjets}) or (\ref{fJjets}).

\subsection{Event shapes}
Shape variables that describe the ``jettiness'' of the final
state have proved useful in $\ee$ annihilation studies;
one of the earliest and still most commonly-used
is the {\em thrust, T}, defined as~\cite{farhi}
\beq\label{Tdef}
T = \max\frac{\sum_i |\vec p_i\cdot\vec n|}{\sum_i |\vec p_i|}
\eeq
where the sum is over all final-state particles and the maximum
is with respect to the direction of the unit vector $\vec n$.
The thrust has value one-half for a fully isotropic final state,
and its value approaches unity as the configuration in the
hadronic centre-of-mass frame becomes more two-jet-like.

If the definition (\ref{Tdef}) is taken over directly
to describe DIS, the value of the thrust is strongly affected by
the properties of the proton remnant.  This is not very satisfactory
since the remnant is not involved in the hard scattering and in any
case is often not seen in the detector.  An alternative definition
involves looking at the final state in the {\em Breit frame} of
reference instead of the hadronic c.m.\ frame.  The Breit frame
is the one in which the momentum transfer from the lepton is
purely spacelike and lies along the $z$-axis: say,
$q^\mu = (0,0,0,Q)$.  In this frame the current jet is usually
in the same hemisphere as $q^\mu$ (the current hemisphere),
while the remnant jet is in the opposite (remnant) hemisphere.
We may therefore define the ``current jet thrust" $T_c$
as $2P/Q$ where $P$ is the total longitudinal momentum
in the current hemisphere in the Breit frame.

The distribution of the current jet thrust has not yet been
calculated to next-to-leading order. A calculation and
experimental measurements of this quantity would be valuable
for $\as$ determination and for comparison with $\ee$ results.
We shall see in the following subsection that recent ideas about
non-perturbative contributions could be tested by such a
comparison.

One shape variable whose distribution can already be predicted
to next-to-leading order is $y_{2+1}$, the value of the jet
resolution $\yc$ at which two jets plus the remnant jet
can just be resolved in the final state. This distribution
is essentially just the derivative of the $1+1$-jet rate
with respect to $\yc$, which can be deduced (for the modified
JADE jet algorithm) from the jet rate calculations described
above.  So far, the data have not been presented in this form.
Again, we shall see below that ideas about non-perturbative
contributions could be tested using this quantity,
by studying the $Q^2$-dependence of the discrepancy
between the perturbative prediction and experiment.

\subsection{$1/Q$ corrections}
A field of recent theoretical activity that needs more
experimental input, which could be provided by studies
of final states in DIS, involves the investigation of
$1/Q$ corrections to hadronic event shapes
\cite{hadro,sterman,manwise,DW,AZ,NS,BB,DMW,CDW}.

In $\ee$ annihilation, it is well known that
many event shape variables receive significant
non-perturbative contributions of the form
$\lambda/Q$, where $\lambda$ is typically of the order
of 1 GeV and $Q$ is the hard process scale, which in
$\ee$ annihilation is the centre-of-mass energy, $Q=\Ecm$.
This is seen for example in the mean value of
the thrust (\fref{meant}): the discrepancy
between the data and the perturbative prediction
(dashed) shows a clear $1/\Ecm$ dependence.  As
also illustrated in \fref{meant},  the full
dependence on $\Ecm$ is well described by the Monte
Carlo event generators {\small JETSET} \cite{jetset}
and \HW\ \cite{herwig}.  The discrepancy
has therefore customarily been described as a
``hadronization correction", estimated according
the the models of the hadronization process that
are built into those programs.  Similar corrections,
also with a $1/Q$ dependence, are found in
differential event shape distributions.

\ffig{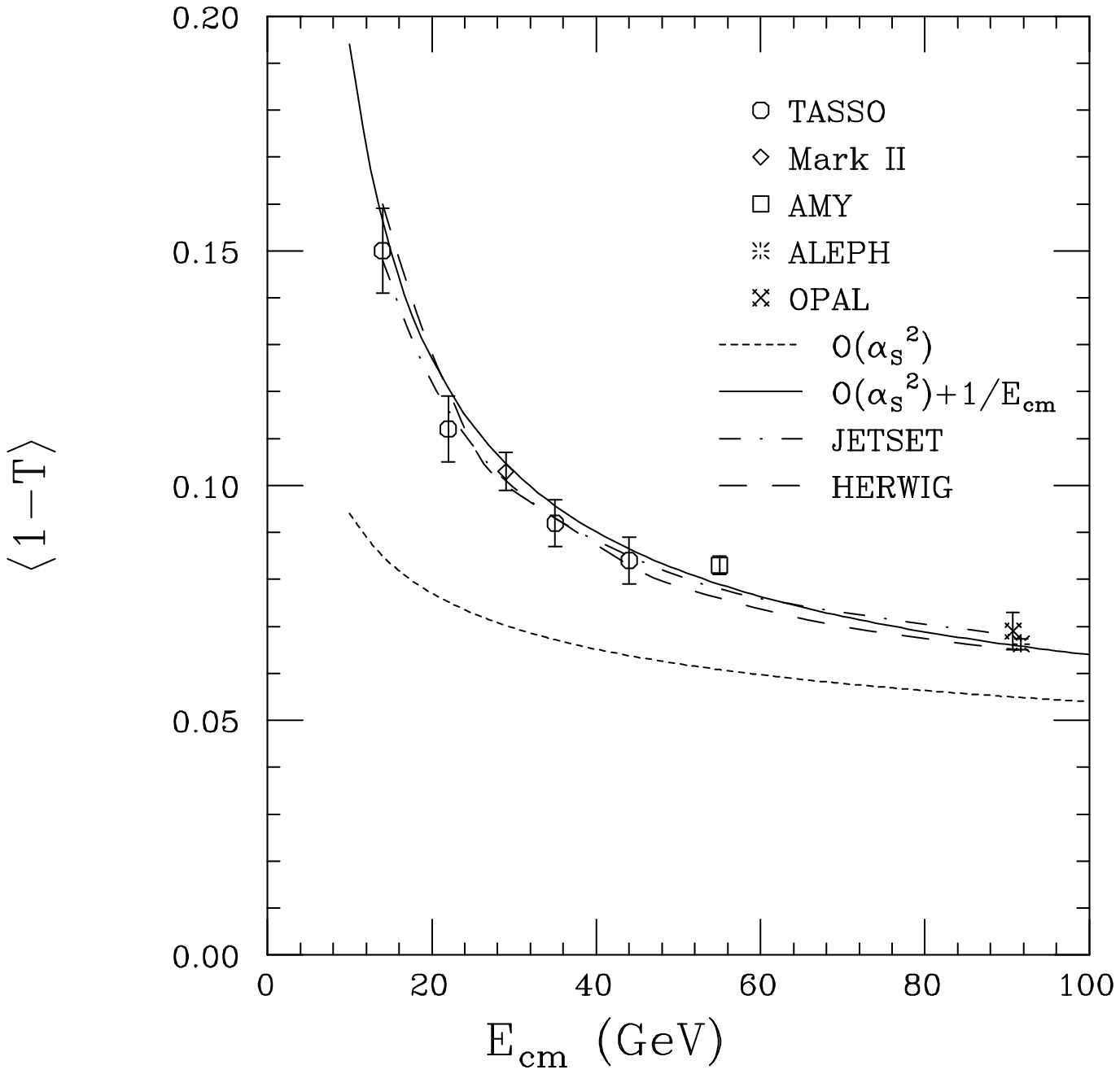}{80mm}{\em Mean value of $1-T$ in $\ee$
annihilation.}{meant}

Recent theoretical studies suggest that $1/Q$
corrections are not necessarily related to
hadronization, but may instead be a universal
soft gluon phenomenon associated with the
behaviour of the running coupling at small
momentum scales~\cite{DW,AZ}. The term `universal'
means that they could be expressible in terms of a
few non-perturbative parameters that are not
themselves calculable but have calculable
process-dependent coefficients.

Final states in deep inelastic scattering are an
excellent place to test these ideas experimentally.
If the conjectured universality is true, there
should be $1/Q$ corrections to shape variables
for DIS final states, where $Q$ is now the
usual DIS momentum transfer variable.
Furthermore the coefficients of $1/Q$ should be
related to those measured in $\ee$
annihilation~\cite{CDW}.

{}From the experimental point of view, DIS
appears to be a good process in which to
study $1/Q$ effects, since the wide range of
$Q^2$ values available in a single experiment
should make it straightforward to disentangle
the relevant power-behaved terms from the
logarithmically-varying perturbative terms. Note
also that the terms in question are dominant
over the more familiar higher-twist corrections,
which behave like $1/Q^2$.

Before giving some illustrative predictions for
DIS final states, let us recall the mechanism proposed
in ref.~\cite{DW} as the source of $1/Q$ corrections.
For definiteness, consider again the mean value of
the thrust in $\ee$ annihilation. The ``improved"
leading-order perturbative prediction of this
quantity is of the form
\beq\label{vev1T}
\VEV{1-T}\sim \int_0^Q d\kt \as(\kt) f_T(\kt,Q)\;.
\eeq
By ``improved" we mean that in the perturbative
prediction we use the running coupling constant
evaluated at a scale given by the transverse
momentum $\kt$ of the emitted final-state
gluon~\cite{askt}. The function $f_T$ turns
out to have the behaviour
\beq
f_T(\kt,Q)\simeq 4C_F/\pi Q\;\;\;\;\mbox{for}\;\kt\ll Q\,.
\eeq
Substituting this and the perturbative expansion for $\as(\kt)$,
\beq\label{alPT}
\as^{\pt}(\kt) = \as(Q)+b\ln\left(\frac{Q}{\kt}\right)
\as^2(Q)+\cdots
\eeq
($b=[33-2N_f]/6\pi$) into eq.~(\ref{vev1T}), we find
\beq
\VEV{1-T}\sim \frac{4C_F}{\pi}\as(Q)\sum_n n!\,[b\as(Q)]^n\;.
\eeq
Because of the $n!\,$ in the coefficient, the series is
strongly divergent.  Therefore by ``improving" the
perturbative prediction we have made it meaningless!

The reason for the divergence is that the perturbative expression
for $\as(\kt)$ diverges at the Landau pole, $\kt=\Lambda$.
Therefore, if we use $\as^{\pt}$ in eq.~(\ref{vev1T}),
the integral is not well defined. On the other hand if we
truncate the series (\ref{alPT}) at any finite
$n$ there is only an integrable divergence at $\kt=0$.
Consequently, for consistency, the series must diverge.

\ffig{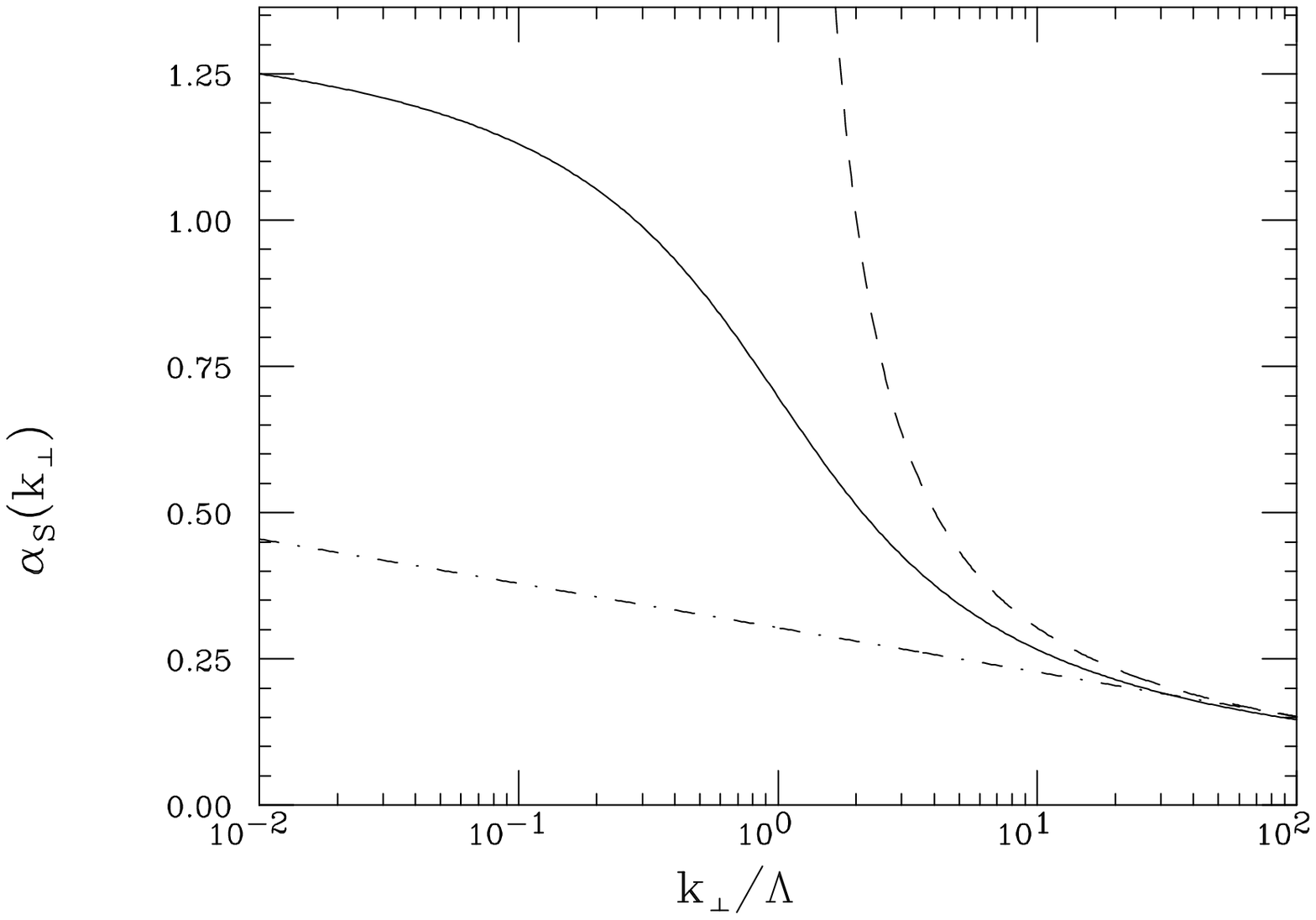}{80mm}{\em Solid curve: possible infrared-finite
behaviour of the strong coupling at low scales. Dashed: one-loop
running $\as(\kt)$. Dot-dashed: expansion of one-loop $\as(\kt)$
to second order in $\as(Q)$ for $Q/\Lambda=100$.}{alfig}

One can attempt to give a meaning to the integral
involving $\as^{\pt}$ by specifying how to deal with
the Landau pole. The various options differ by
terms of order $1/Q$ and therefore one is led to
expect a term of this form in the true answer,
whose coefficient cannot be determined within
perturbation theory. This is an example of the
(infrared) {\em renormalon ambiguity}
\cite{sterman,NS,BB,mueller}.

In ref.~\cite{DW} a more phenomenological approach
is proposed. Suppose that a meaningful, universal
form of $\as(\kt)$ can be defined for all values of
$\kt$; for example, the solid curve in \fref{alfig}.
Then the integral
\beq\label{abarmu}
\int_0^\mu d\kt \as(\kt)\equiv \mu\,\a0(\mu)
\eeq
exists for all $\mu\geq 0$. For $\mu$ sufficiently
large ($\mu\gg\Lambda$) we have
$\as(\kt)\simeq \as^{\pt}(\kt)$ for $\kt>\mu$.
Therefore we may write
\beq
\VEV{1-T} \simeq \VEV{1-T}^{\pt} + \VEV{1-T}^{\pw}
\eeq
where $\VEV{1-T}^{\pt}$ represents the second-order
perturbative prediction, while (for $\Lambda\ll\mu\ll Q$)
\beeq\label{Tpow}
\VEV{1-T}^\pw &=& \frac{4C_F}{\pi Q}
\int_0^\mu d\kt \Biggl[\as(\kt)\nonumber \\
& &-\as(Q)-b\ln\left(\frac{Q}{\kt}\right)
\as^2(Q)\Biggr]\nonumber \\
&=& -\frac{4C_F\mu}{\pi Q}
 \Biggl[\a0(\mu)-\as(Q)\nonumber \\
& &-b\left(\ln\frac{Q}{\mu}+1\right)
\as^2(Q)\Biggr]\;.
\eeeq
The negative terms in the integrand express the fact
that we have subtracted the perturbative contribution
from $\kt<\mu$ and replaced it by the non-perturbative
expression (\ref{abarmu}).  As shown in \fref{alfig},
the subtracted part is probably smaller than
$\as(\kt)$ and therefore we expect a positive power correction.
Good agreement with the thrust data shown in \fref{meant}
is found for $\a0(2\;\mbox{GeV}) = 0.52\pm 0.04$~\cite{DW}.

Using the same approach, one finds power corrections like (\ref{Tpow})
for a wide range of final-state observables in hard processes, including
deep inelastic scattering~\cite{DMW,CDW}. In general the correction is of the
form\footnote{In ref.~\cite{DW} there is a constant added to the final
$1/r$, corresponding to the use of a different renormalization scheme
for the definition of $\ap$.}
\beeq\label{Fpow}
F^\pw &=& a\frac{4C_i}{\pi r}\left(\frac{\mu}{Q}\right)^r
\log^s\left(\frac{Q}{\mu}\right)
\Biggl[\ap(\mu) \nonumber \\
& &- \as(Q) -
b\left(\ln\frac{Q}{\mu}+\frac{1}{r}\right)\as^2(Q)\Biggr]
\eeeq
where $a$, $r$ and $s$ ($s=0$ or 1) are constants that depend on the
observable $F$, $C_i$ is the relevant colour factor ($C_i=C_F=4/3$
for the quantities we are considering) and
\beq\label{apdef}
\ap(\mu) \equiv r \mu^{-r} \int_0^\mu d\kt \kt^{r-1} \as(\kt)\;.
\eeq
Predictions for the coefficient $a$ and the exponents $r$ and $s$ for
various $\ee$, deep inelastic and Drell-Yan observables are listed in
table~1 \cite{CDW}.

\begin{table}
\begin{center}\begin{tabular}{cccc}\hline
$F$ & $a$ & $r$ & $s$ \\ \hline
 & & & \\
$\VEV{T}_\see$ & --1 & 1 & 0 \\
 & & &\\
$\VEV{C}_\see$ & $3\pi/2$ & 1 & 0 \\
 & & &\\
$(\sL/\st)_\see$ & $\pi/4$ & 1 & 0 \\
 & & &\\
$\VEV{M_j^2/Q^2}_\see$ & 1 & 1 & 0 \\
 & & &\\
$\VEV{B}_\see$ & 1 & 1 & 1 \\
 & & &\\
$\VEV{y_3}^{\mbox{\scriptsize J}}_\see$ & 1 & 1 & 0 \\
 & & &\\
$\VEV{y_3}^{\kt}_\see$ & - & 2 & - \\
 & & &\\ \hline
 & & &\\
$\VEV{T_c}_\sds$ & --1 & 1 & 0 \\
 & & &\\
$\VEV{y_{2+1}}^{\mbox{\scriptsize mJ}}_\sds$ &
$\sqrt{\frac{x}{1-x}}$ & 1 & 0 \\
 & & &\\
$\VEV{y_{2+1}}^{\mbox{\scriptsize fJ}}_\sds$ & 1 & 1 & 0 \\
 & & &\\
$\VEV{y_{2+1}}^{\kt}_\sds$ & - & 2 & - \\
 & & &\\ \hline
 & & &\\
$\VEV{q_t/Q}_\sdy$ & 1 & 1 & 1 \\
 & & &\\
$\VEV{q_t^2/Q^2}_\sdy$ & - & 2 & - \\
 & & &\\
\hline \end{tabular}
\caption{\em Predictions for power corrections to $\ee$,
deep inelastic and Drell-Yan observables.}
\vspace*{-0.25cm}
\end{center}
\end{table}

The $\ee$ quantities $T$, $C$ and $\sL$ are the thrust, $C$-parameter
and longitudinal cross section, discussed in ref.~\cite{DW}.
$M_j^2$ represents the jet mass-squared.
In one-loop order this may correspond to the heavy jet mass-squared
$M^2_h$ or the heavy-light jet difference $M^2_h-M^2_l$,
there being no difference between these quantities until
multi-jet contributions become significant.
Discrepancies between the coefficients of $1/Q$ corrections
to these quantities therefore measure the non-perturbative
effect of multi-jet contributions not associated with the
running of $\as$ \cite{NS}. In our treatment the effective
expansion parameter here is $\a0\sim 0.5$, and so the predicted
coefficients would not be expected to be more reliable than
about $\pm 50\%$. Similarly for $B$, which represents either
the total or wide jet broadening~\cite{broad} at one-loop order.

Quantities like $\VEV{B}_\see$ have enhanced ($s=1$) leading power
corrections associated with the logarithmically growing phase
space for gluon emission with limited $\kt$.  Such quantities
also have non-enhanced terms with the same power, but these
are more difficult to predict because they receive
contributions from the non-soft, wide-angle region.

The quantities $\VEV{y_3}^{\mbox{\scriptsize J}}_\see$
and $\VEV{y_3}^{\kt}_\see$ are the values of the
jet resolution $\yc$ at which three jets are just resolvable,
using the JADE and $\kt$-algorithms. The $\kt$-algorithm
is ``better" in that its correction is suppressed by one
extra power of $Q$. This could be related
to the observation that this algorithm has
smaller hadronization corrections.

Note that predictions for the coefficient $a$ and the logarithmic
exponent $s$ are not shown in table~1 for observables which, like
$\VEV{y_3}^{\kt}_\see$, have no $1/Q$ correction. When the leading
power correction is of order $1/Q^2$, there are other mechanisms
that can give contributions of the same order, making the
predictions less straightforward~\cite{BB,DMW}.

Turning to the observables for deep inelastic scattering,
we consider first the current jet thrust as defined previously.
For this quantity the power correction (unlike the perturbative
contribution) is predicted to be approximately equal to
that for the thrust in $\ee$ annihilation at $\Ecm = Q$,
independent of Bjorken $x$. As discussed above,
corrections to this relationship are higher-order
in $\a0$, and so the equality should be good to
about $\pm 50\%$.

The observable $\VEV{y_{2+1}}^{\mbox{\scriptsize A}}_\sds$ is the
mean jet resolution at which two jets plus the remnant jet
are just resolvable in DIS using jet clustering algorithm A.
We denote by A = mJ, fJ and $\kt$ the modified and factorizable
JADE and $\kt$ algorithms defined above. The mJ algorithm is the
only one for which next-to-leading order predictions are available
at present~\cite{nlo}. Note that in this case the factor of
$\sqrt{x/(1-x)}$ in the coefficient $a$ means that the power correction
should be approximately the same as that in $\ee$ annihilation at
$\Ecm = W = Q\sqrt{(1-x)/x}$, rather than at $\Ecm = Q$.

In the factorizable JADE-type algorithm (fJ), on the other hand,
the scale of the power correction is $Q$ instead of $W$, making
it larger than that for the modified JADE algorithm. One should
remember, however, that this is also true of
the perturbative terms, so the relative correction is similar.

As in $\ee$ annihilation, the $\kt$-type algorithm for
DIS~\cite{DISkt} has a smaller (order $1/Q^2$) correction,
suggesting a smaller ``hadronization" correction.

For Drell-Yan processes, $q_t$ in table~1 is the transverse momentum
of the lepton pair and $Q$ is its invariant mass.  The mean value
of $q_t/Q$ is expected to have a $1/Q$ correction, again
log-enhanced due to the growing phase space for gluon emission.
The possibility of a $1/Q$ power correction to the Drell-Yan
cross section itself has also been considered~\cite{sterman} but
seems less likely~\cite{BB,DMW}.

\subsection{Fragmentation studies}
In addition to jet rates and event shapes, one can investigate the
fragmentation properties of jets in DIS final states~\cite{H1frag,Zfrag}.
Here again it is advantageous to study the current hemisphere in the
Breit frame, which should be comparable to a single hemisphere of an
$\ee$ final state at $\Ecm = Q$.  In particular the average multiplicity
of charged hadrons should be asymptotically independent of Bjorken
$x$, with a $Q^2$-dependence given by the next-to-leading-logarithmic
(NLL) prediction~\cite{mult}
\beq\label{nchqcd}
\VEV{n_{ch}} \sim A \as^b(Q) \exp[c/\sqrt{\as(Q)}]\;,
\eeq
where the normalization $A$ is not predicted (but should be equal
to that for a single $\ee$ hemisphere) while the exponents
$b$ and $c$ are known constants. As shown in \fref{nchavfig},
the data on charged multiplicity
in current fragmentation do follow the $Q$-dependence of the $\ee$
data, which in fact agree well with the QCD prediction (\ref{nchqcd}).

\ffig{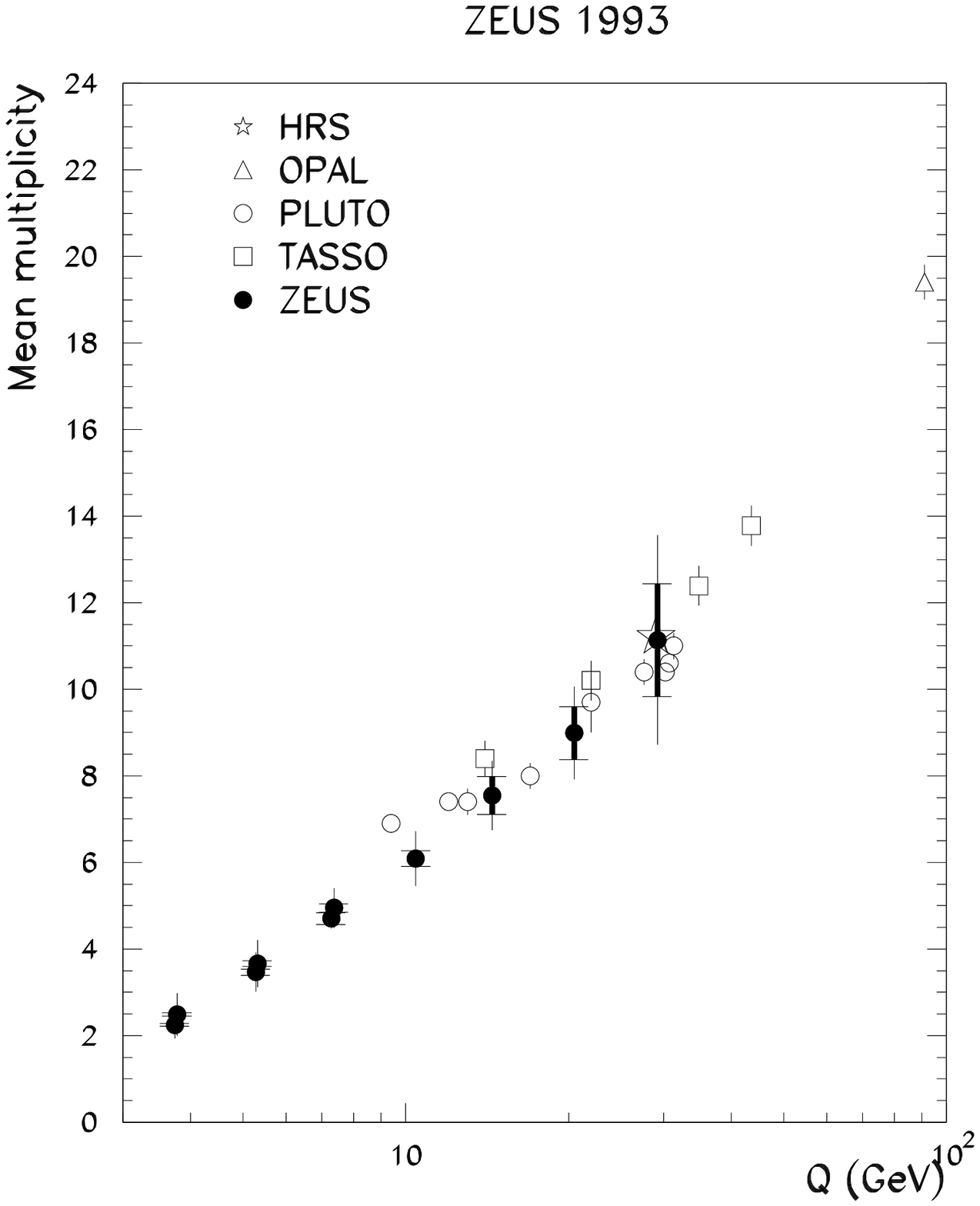}{80mm}{\em Mean charged multiplicity in $\ee$
(open symbols) and DIS final states.}{nchavfig}

One can go on to compare other properties of the charged multiplicity
distribution in current jet fragmentation with the corresponding results
for a single hemisphere in $\ee$ annihilation.  For example, the
second binomial moment $\VEV{n_{ch}(n_{ch}-1)}$ should satisfy the
$x$-independent relation~\cite{MalWeb}
\beeq\label{R2qcd}
\frac{\VEV{n_{ch}(n_{ch}-1)}}{\VEV{n_{ch}}^2} &\sim& \frac{7}{4}
-\left( \frac{55}{8} -\frac{5}{81} N_f\right)A \sqrt{\frac{\as(Q)}{6\pi}}
\nonumber \\
& & + {\cal O}[\as(Q)]\;.
\eeeq
The data show that the shape of the distribution does appear
to depend on $Q^2$ and not explicitly on $x$, but a comparison
with eq.~(\ref{R2qcd}) has not yet been presented.

Studies of the current jet fragmentation function,
i.e.\ the inclusive single particle momentum distribution, are also
being performed~\cite{H1frag,Zfrag}. One expects that the distribution
of the logarithm of the momentum fraction, $\xi = \ln(1/x_p)$, where
$x_p = 2p/Q$ in the Breit frame current hemisphere, should be
the same as that in $\ee$ annihilation at $\Ecm = Q$.
The distribution in $\xi$ should be approximately Gaussian,
with a peak that moves linearly with $\ln Q$. More precisely,
to NLL accuracy we expect the peak to be at $\xi=\xi^*$ where
\beq\label{xipeak}
\xi^* \sim \frac 1 2 \ln\left(\frac Q {\lef}\right)\left[
1 + \left( \frac{11}{2} -\frac{1}{27} N_f\right)
\sqrt{\frac{\as(Q)}{6\pi}} + \ldots\right]\,.
\eeq
Here $\lef$ is an effective scale (not necessarily equal to $\lms$)
and the ellipsis represents ${\cal O}[\as(Q)]$ corrections.  As shown
in \fref{xpeakfig}, a linear leading $\ln Q$ dependence is
seen, in good agreement with the $\ee$ data and in disagreement
with a simple phase space model.

\ffig{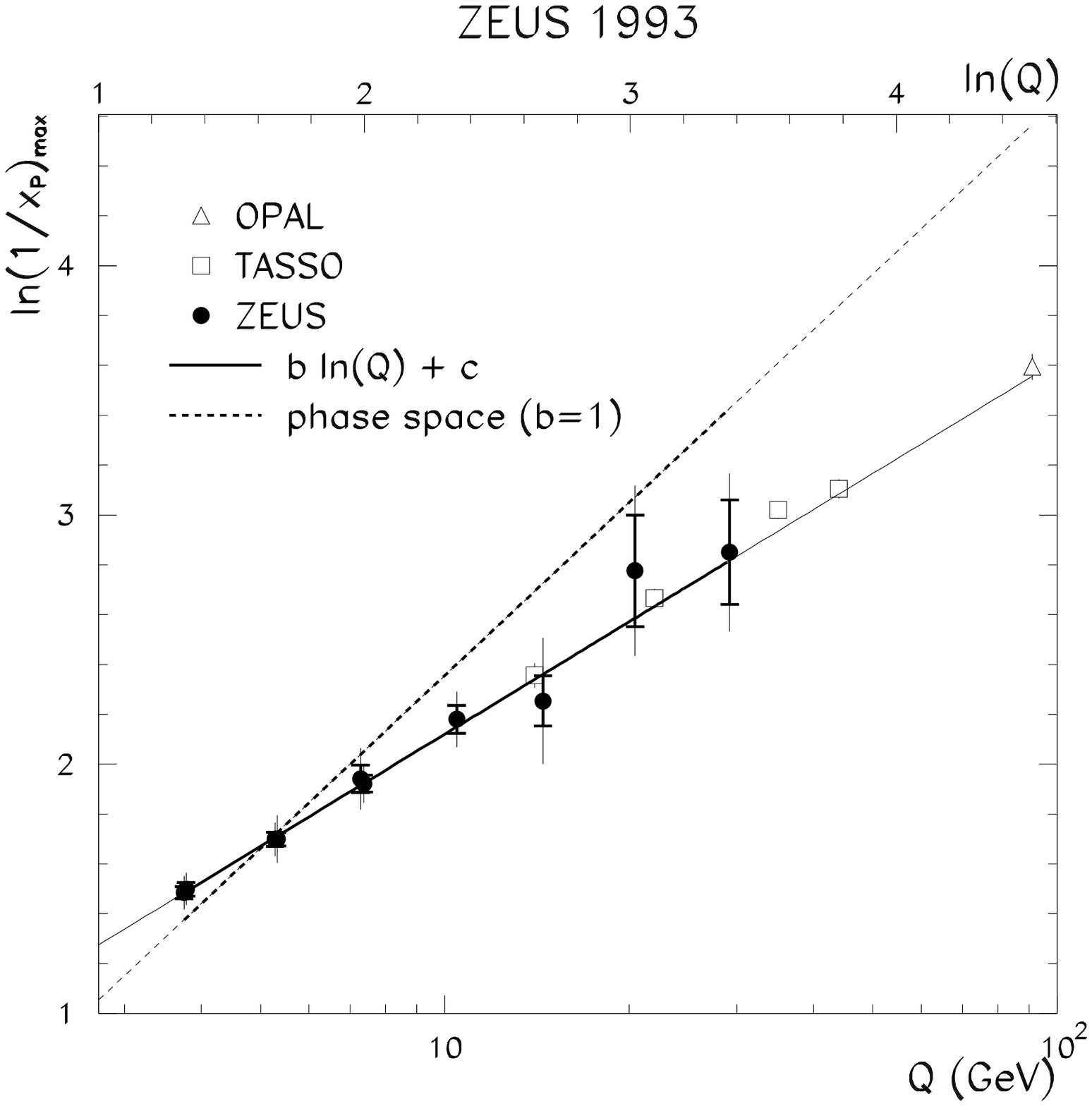}{80mm}{\em Position of the peak in the $\ln(1/x_p)$
distribution for $\ee$ (open symbols) and DIS final states.}{xpeakfig}

Again, it will be interesting to test other QCD predictions
concerning the form of the $\xi$-distribution. For example,
the NLL result for the r.m.s.\ width $\sigma$ is~\cite{FW}
\beq\label{xisig}
\sigma^2 \sim \frac 1 {36} \ln\left(\frac Q {\lef}\right)\left[
\sqrt{\frac{6\pi}{\as(Q)}} - \frac{33}{8} + \frac{1}{4} N_f
+ \ldots\right]\;.
\eeq

As the data become more precise, it will be necessary to go beyond the
NLL treatment, and systematic differences between $\ee$ and DIS
jet fragmentation should become manifest, due to the different higher-order
corrections, quark flavour composition and threshold effects in the two
processes.

\section{Small x Final States}
\subsection{Theoretical remarks}
One of the areas of greatest current interest in DIS physics is
QCD dynamics in the region of small Bjorken $x$. Although HERA
has opened up a much larger window on the small-$x$ region, it
is still the case at present that different theoretical approaches
can account for the increase in the totally inclusive structure
function $F_2$ at small $x$~\cite{stirling}.  It is therefore important
to consider less inclusive aspects of DIS, and in particular to identify
features of the final state that might distinguish between the different
mechanism proposed for the small-$x$ rise in $F_2$.

Outside the small-$x$ region, the standard approach embodied in the
DGLAP $Q^2$-evolution equations~\cite{dglap} has a firm theoretical
basis in the operator product expansion, renormalization group and
factorization theorems.  A simple extrapolation of DGLAP evolution
to small $x$ still appears to be consistent with the structure
function data.  The corresponding final states are dominated by
strongly ordered configurations of transverse momentum $k_t$ and
virtuality $k^2$ along the ladder of partons connecting the soft,
low-virtuality proton constituents to the hard subprocess
(\fref{ladderfig}). Such configurations generate the greatest
number of logarithms of $Q^2$ in the evolution of the structure
functions. Sub-leading corrections are classified according to
the number of $k_t$-disordered rungs in the ladder.
The \HW\ event generator~\cite{herwig},
as well as the matrix element plus parton shower (MEPS) option of
{\small LEPTO} \cite{lepto}, are implementations of DGLAP evolution,
with different approximate treatments of sub-leading terms.
The {\small ARIADNE} generator~\cite{ariadne} uses the colour dipole
model (CDM)~\cite{dipole}, in which disordered transverse momentum
configurations are permitted but the radiated gluons are regarded
as emitted by the proton remnant, so the connection with the
picture in \fref{ladderfig} is not so clear.

\ffig{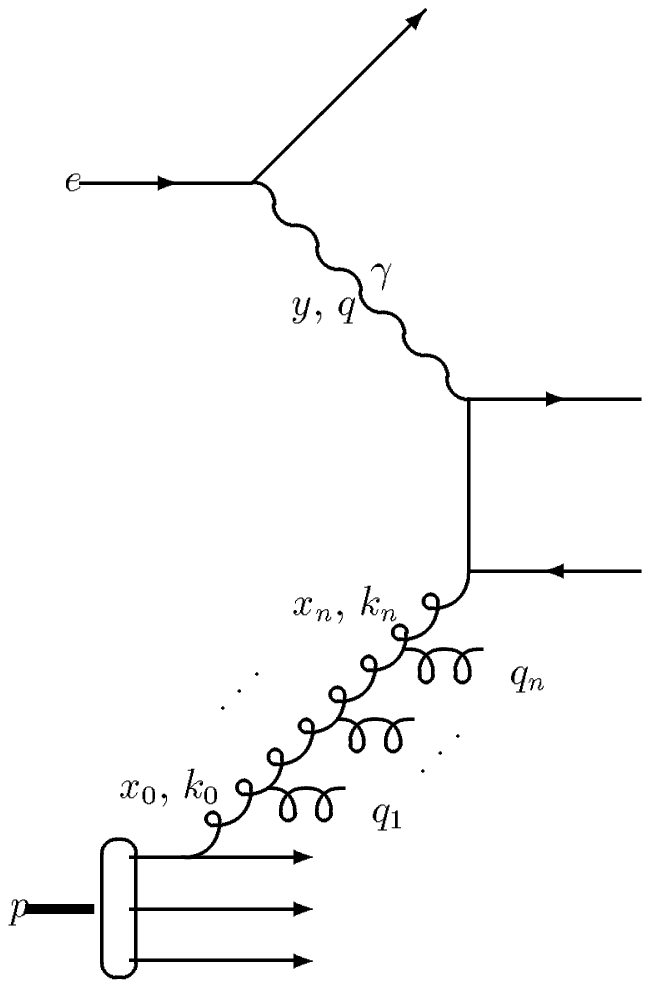}{80mm}{\em Ladder graph for DIS at small $x$.}
{ladderfig}

At very small $x$, a classification of contributions
according to the degree of ordering in $x$ rather than $k_t$
becomes more appropriate, because logarithms of $x$ dominate
over those of $Q^2$. The most widely used approach is that based
on the BFKL equation~\cite{bfkl}, which describes the $x$-dependence
of the structure function in terms of a sum over ladder graphs that
exhibit strong ordering in $x$ without ordering in transverse
momentum. The original BFKL derivation was based on multi-Regge
dynamics, in which $k_t$ values are limited.
More recently, the same equation for the $x$-dependence of the
structure function has been derived by Ciafaloni, Catani,
Fiorani and Marchesini (CCFM)~\cite{ccfm} from an approach based
more closely on QCD, in particular on the coherence properties of
the theory at small $x$. Although the CCFM approach leads to the
same equation as BFKL when summed over all final states, it gives
a different picture of less inclusive quantities~\cite{GM95}.

In order to compare the results of the two approaches, it is
convenient to perform a Mellin transformation, introducing the
moment variable $\omega$ conjugate to $x$:
\beq
\tilde F(\omega) = \int_0^1 dx\,x^\omega F(x)\;.
\eeq
Then the objective at small $x$ is to sum to all orders in $\as$ the
terms that are most singular at $\omega=0$.  In both approaches the
structure function is expressed as a sum over ladder contributions
as in \fref{ladderfig}:
\beq
\tilde F(\omega,Q/\mu) = \sum_{r=0}^{\infty}\tilde F^{(r)}(\omega,Q/\mu)
\eeq
where $r$ is the number of rungs in the ladder and $\mu$ is an infrared
cutoff.  In the BFKL case one may write the ladder contributions in
the form
\beq\label{FrBFKL}
\tilde F^{(r)}_{\BFKL} \sim \int_{q_i=\mu}^Q \prod_{i=1}^r
\frac{d^2q_i}{\pi q_i^2}\frac{dz_i}{z_i} z_i^\omega \bar\as
\DelR(z_i,k_{ti})\;,
\eeq
where the variable $q_i$ is related to the transverse momentum
$q_{ti}$ and longitudinal momentum fraction $(1-z_i)$ of an emitted
gluon, $q_i = q_{ti}/(1-z_i)$, $k_{ti} = \sum _{j=1}^{i} q_{tj}$
is the corresponding exchanged transverse momentum, and
$\bar\as = C_A\as/\pi$. The function $\DelR(z_i,k_{ti})$
is a Regge form factor,
\beq\label{DelRegge}
\DelR(z_i,k_{ti}) = \exp\left(-\bar\as\int_{z_i}^1 \frac{dz}{z}
\int_{\mu^2}^{k_{ti}^2} \frac{dk_t^2}{k_t^2}\right)\;.
\eeq

The CCFM expression corresponding to eq.~(\ref{FrBFKL}) is
\beeq\label{FrCCFM}
\tilde F^{(r)}_{\CCFM} &\sim & \int_{q_i=0}^Q \prod_{i=1}^r
\frac{d^2q_i}{\pi q_i^2}\frac{dz_i}{z_i} z_i^\omega \bar\as
\DelN(z_i,k_{ti},q_i)\nonumber \\
& & \times\Theta(q_i-z_{i-1}q_{i-1})\Theta(q_1-\mu)\;.
\eeeq
The main differences are the ordering condition $q_i>z_{i-1}q_{i-1}$,
which corresponds to {\em angular ordering} of gluon emission due
to coherence, and the replacement of the Regge form factor
(\ref{DelRegge}) by the `non-Sudakov' form factor
\beq\label{DelNS}
\DelN(z_i,k_{ti},q_i) = \exp\left(-\bar\as\int_{z_i}^1 \frac{dz}{z}
\int_{z^2q_i^2}^{k_{ti}^2} \frac{dk_t^2}{k_t^2}\right)\;.
\eeq
This leads to the same asymptotic behaviour for the structure function:
\beq
\tilde F_{\BFKL} \sim \tilde F_{\CCFM} \sim
1 + 2\bar\as\frac{t}{\omega}+ 2\bar\as^2\frac{t^2}{\omega^2}+\ldots\;,
\eeq
where $t=\ln(Q/\mu)$. However, the final state properties are different.
For example, the mean number of ladder rungs, which gives the mean
number of primary emitted gluon jets, is related to the quantity
$\sum r\tilde F^{(r)}$, for which one finds~\cite{GM95}
\beeq
\sum r \tilde F_{\BFKL}^{(r)} &\sim &
2\bar\as\frac{t}{\omega}+ 6\bar\as^2\frac{t^2}{\omega^2}
+ \ldots\nonumber \\
\sum r \tilde F_{\CCFM}^{(r)} &\sim &
2\bar\as\frac{t}{\omega}+ 4\bar\as^2\left(\frac{t^2}{\omega^2}
+\frac{t}{\omega^3}\right) + \ldots\;.
\eeeq
Notice that the BFKL series for this quantity, like that for the
structure function itself, involves at most one extra factor of
$1/\omega$ for each power of $\as$, whereas the CCFM series contains up
to two such factors.  The new terms are subleading with respect to the
number of logarithms of $Q$ but dominate at small $x$ ($\omega\to 0$).
The corresponding expansions for the mean number of primary emitted jets
are
\beeq
\VEV{r}_{\BFKL} &\sim &
2\frac{\bar\as}{x}\ln\frac Q \mu
\left[1+\bar\as\ln\frac 1 x\ln\frac Q \mu + {\cal O}(\as^3)\right]
\;, \nonumber \\
\VEV{r}_{\CCFM} &\sim &
2\frac{\bar\as}{x}\ln\frac Q \mu
\Bigl[1+\bar\as\ln^2\frac 1 x -\frac 1 6\bar\as^2\ln^4\frac 1 x
\nonumber \\
& & + {\cal O}(\as^3)\Bigr]\;.
\eeeq
We see that more emission at small $x$ is predicted by the CCFM
treatment. The reason is that the angular ordering prescription
in eq.~(\ref{FrCCFM}) corresponds to an increased phase space
compared to that available in eq.~(\ref{FrBFKL}) for any fixed
value of the infrared cutoff $\mu$.

So far, the CCFM approach has not reached the stage of detailed
phenomenological application, and so the quantitative differences between
the CCFM and BFKL predictions for final states in the accessible domain
of $x$ and $Q^2$ are not yet clear.  Predictions of CCFM and DGLAP
evolution were compared in ref.~\cite{MWsmallx} at large $Q^2$
down to $x\sim 10^{-5}$ and the differences were not great;
one would expect CCFM and BFKL to be even more similar.

\subsection{Transverse energy flow}
The new basic feature of small-$x$ physics (in both the BFKL and CCFM
formulations) is the loss of strong ordering of exchanged and emitted
transverse momenta. This suggests that one should see extra transverse
energy emission in small-$x$ events, especially in the region well
separated from both the current and remnant jets.  The HERA experiments
do see a substantial transverse energy per unit rapidity in this region
(about 2 GeV per unit rapidity) \cite{H1Et,H1fwd,haas}, with some indication of
a rise at small $x$, as shown for example by the H1 data in \fref{Etxfig}.
The curves show that standard DGLAP evolution as implemented in
the MEPS model (defined above) does not produce such a rise, while
the colour dipole model and a BFKL-based calculation~\cite{durham} do,
presumably because of their inclusion of $k_t$-disordered contributions.

\ffig{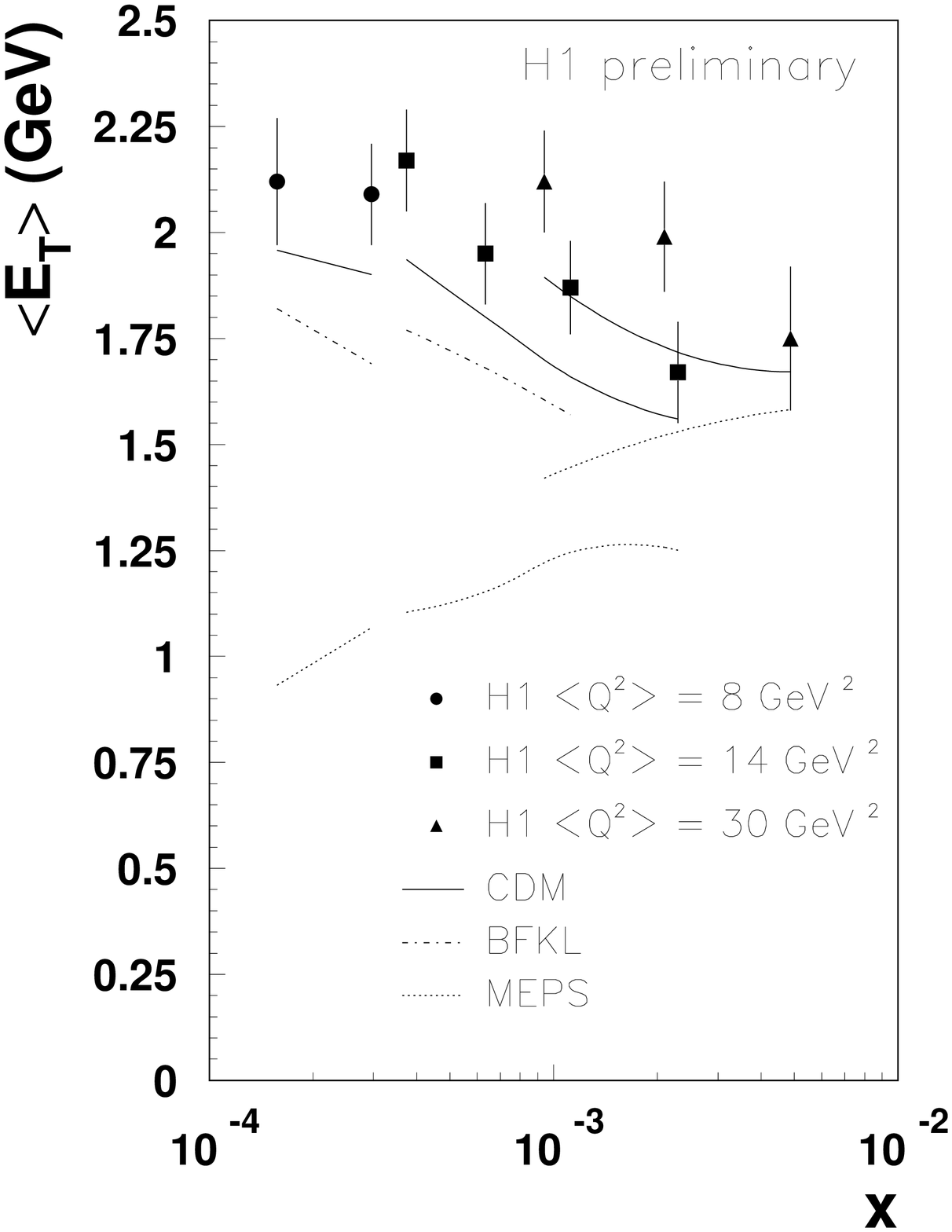}{80mm}{\em Mean transverse energy flow per unit rapidity
(at zero rapidity in the hadronic c.m.\ frame), as a function of Bjorken
$x$, for various ranges of $Q^2$.}{Etxfig}

One point to bear in mind when comparing theoretical predictions for
transverse energy flow is that they are highly sensitive to the correct
treatment of the hard scattering process that takes place at the top of
the ladder in \fref{ladderfig}.  This is because the mean
emitted transverse energy is obtained by inserting a factor of $k_{ti}$
into the integration in eq.~(\ref{FrBFKL}) or (\ref{FrCCFM}).  As a
result, the integral becomes more sensitive to the region $k_{ti}\sim Q$.
In the case of the DGLAP evolution equation, which assumes strong
ordering in $k_t$, the $k_t$-ordering above the $i$-th rung is in fact
destroyed by inserting a factor of $k_{ti}$, and so the relevance of
the equation becomes questionable.  Preferably, one should compute the
transverse energy emission at a particular rapidity from the full hard
scattering matrix elements for producing additional partons in that
direction.

An illustration of the above point was provided by the
\HW\ (version 5.7 and earlier) prediction for the transverse
energy flow, which had a deficit in the rapidity range
of interest.
This was because the parton shower approximation to the
hard matrix elements (with the choice of evolution variables
used in \HW) left a region of phase space unpopulated by emitted
partons.  In version 5.8 a matrix element correction was included
and the transverse energy flow became consistent with the data,
without any retuning of parameters~\cite{seymour}.

\subsection{Associated forward jets}
The most reliable indicator of new physics at small $x$ still seems to be
the associated production of jets in the forward (proton) direction with
relatively large momentum fractions $x_\jet$ and transverse momenta
$q_{t\jet}\sim Q$ \cite{muelfwd}.  In this region there is no phase space
for DGLAP evolution with transverse momentum ordering, whereas the scope for
BFKL evolution in $x$ remains large.  Thus a strong increase in production of
such jets is expected with decreasing Bjorken $x$ \cite{KMS}.  Since the cross
section is fully inclusive with respect to what lies between the forward and
current jets, the CCFM approach leads to the same conclusion.

\ffig{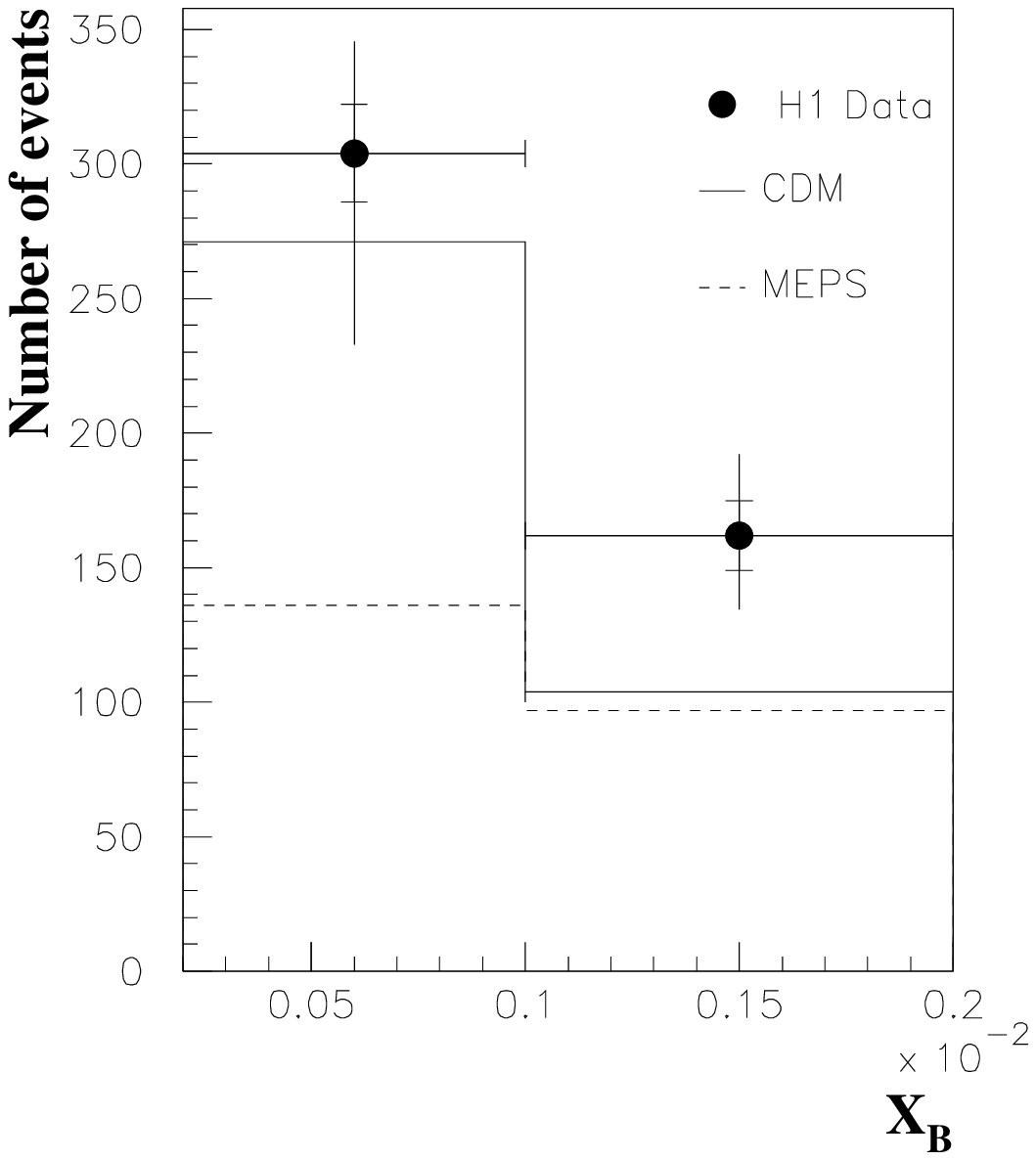}{80mm}{\em Forward jet production rate as a function of
Bjorken $x$.}{forjetfig}

The H1 data (\fref{forjetfig}) \cite{H1fwd} do indeed show a high event rate
at small $x$.  The data shown are for events at $Q^2\sim 20$ GeV$^2$ with
jets with $x_\jet >0.025$, $0.5< q_{t\jet}^2/Q^2<4$ and $q_{t\jet}> 5$
GeV.  The predictions of the colour dipole and MEPS models are also shown.
While the colour dipole model does show a rise at small $x$, it predicts an
even lower rate than MEPS if the cut on $x_\jet$ is increased to
0.05 \cite{H1fwd}.

At present the lack of a full BFKL-based event generator prevents a
comparison with the BFKL prediction including selection cuts and detector
and hadronization corrections. However, a parton-level BFKL calculation
\cite{KMS} predicts a ratio of about 1.6 between the two $x$ bins
in \fref{forjetfig}, which is consistent with the data.

\section{Exotic QCD}
\subsection{Instanton-induced processes}
An intriguing possibility raised in recent theoretical papers
\cite{bb,rs,grs}, but not so far subjected to experimental
investigation, is that fundamentally non-perturbative features of QCD may
lead to exotic new processes in DIS.  Such processes can arise from the
structure of the QCD vacuum, which contains topologically non-trivial
gauge field configurations characterized by an integer value of a
quantity called the Chern-Simons number, $\NCS$.  Configurations with
adjacent values of $\NCS$ are separated by an energy barrier. The
process of tunnelling through this barrier is represented by a space-time
configuration called an {\em instanton}~\cite{instantons}.

An important feature of the instanton tunnelling process is that the
change in the gauge field induces a change in the fermion sector of the
theory.  In the case of QCD the fermionic change involves the axial
charge $Q_5$:
\beq\label{delQ5}
\Delta Q_5 = 2\Delta \NCS = 2 N_f
\eeq
where $N_f$ is the number of active flavours. Since the axial charge
measures the chirality or handedness of the fermions, this implies that
the process is chirality-violating.  In electroweak interactions the
effect is more spectacular: the tunnelling process leads to violation of
baryon- and lepton-number conservation \cite{instantons}. Thus it is possible
(though at present considered unlikely) that instanton-induced processes
could be responsible for the net baryon number of the universe. This makes
it all the more interesting to look for the analogous processes in QCD, which
turn out to have much larger predicted cross sections, essential because
$\as\sim 1/8$ is much larger than $\alpha_w \sim 1/30$. (For a tunnelling
process, a factor of $1/\alpha$ appears in the exponent of the transition
rate - see eq.~(\ref{instcs}) below.)

Taking $N_f=3$, eq.~(\ref{delQ5}) implies that processes like
\beq
\gamma^*\to u_L \bar u_R\,,\;\; \bar u_R + g
\to \bar u_L u_L d_L s_L \bar u_L\bar d_L\bar s_L + \mbox{gluons}
\eeq
can occur (\fref{instantfig}). The subscripts indicate that the
outgoing quarks and antiquarks are all left-handed.  The chirality
violation would be difficult to observe experimentally, but the
relatively high multiplicity, including excess strange (possibly
even charmed) particle production, should be characteristic.  The typical
number of gluons produced is expected to be of order $ \pi/2\as\sim 10$,
further enhancing the multiplicity.  Another distinctive feature is the
relative isotropy of the final state in the rest-frame of the instanton
subprocess, which transforms into a band about two units wide in the
lab rapidity distribution, in contrast to the more normal jet-like
configurations, as indicated in the lower half of \fref{instantfig}.

\ffig{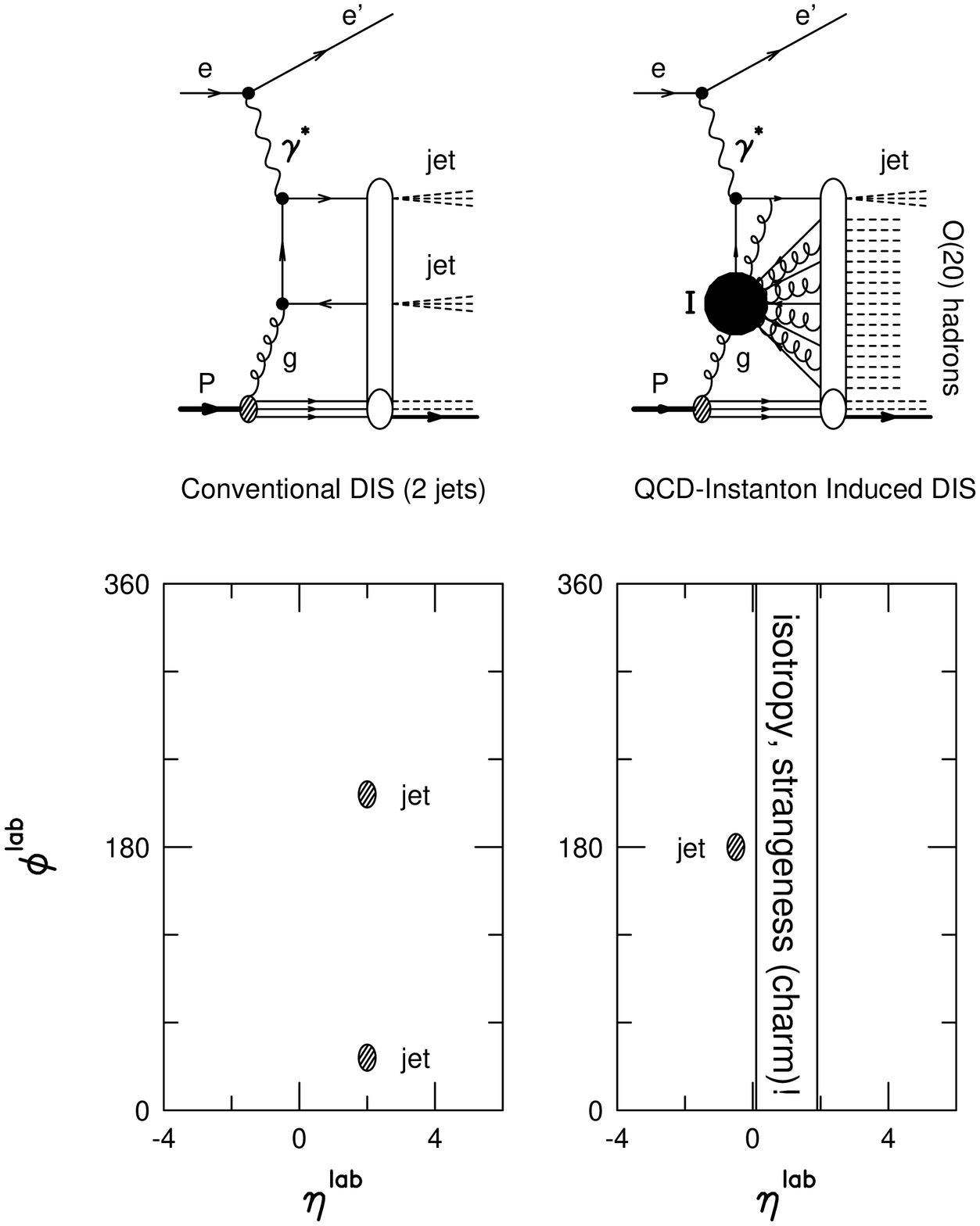}{80mm}{\em Comparison between conventional and
instanton-induced DIS events.}{instantfig}

A Monte Carlo generator for such events is under
development~\cite{grsMC}: a typical simulated event is
depicted in \fref{legofig}.  The isolated particle at
$\eta_{\lab}\simeq -2.5$ is the electron, while the
current-quark jet is around $\eta_{\lab}\simeq -0.5$.
The densely populated  band at $\eta_{\lab}\simeq 2-4$
is the final state from the instanton-induced process.

\ffig{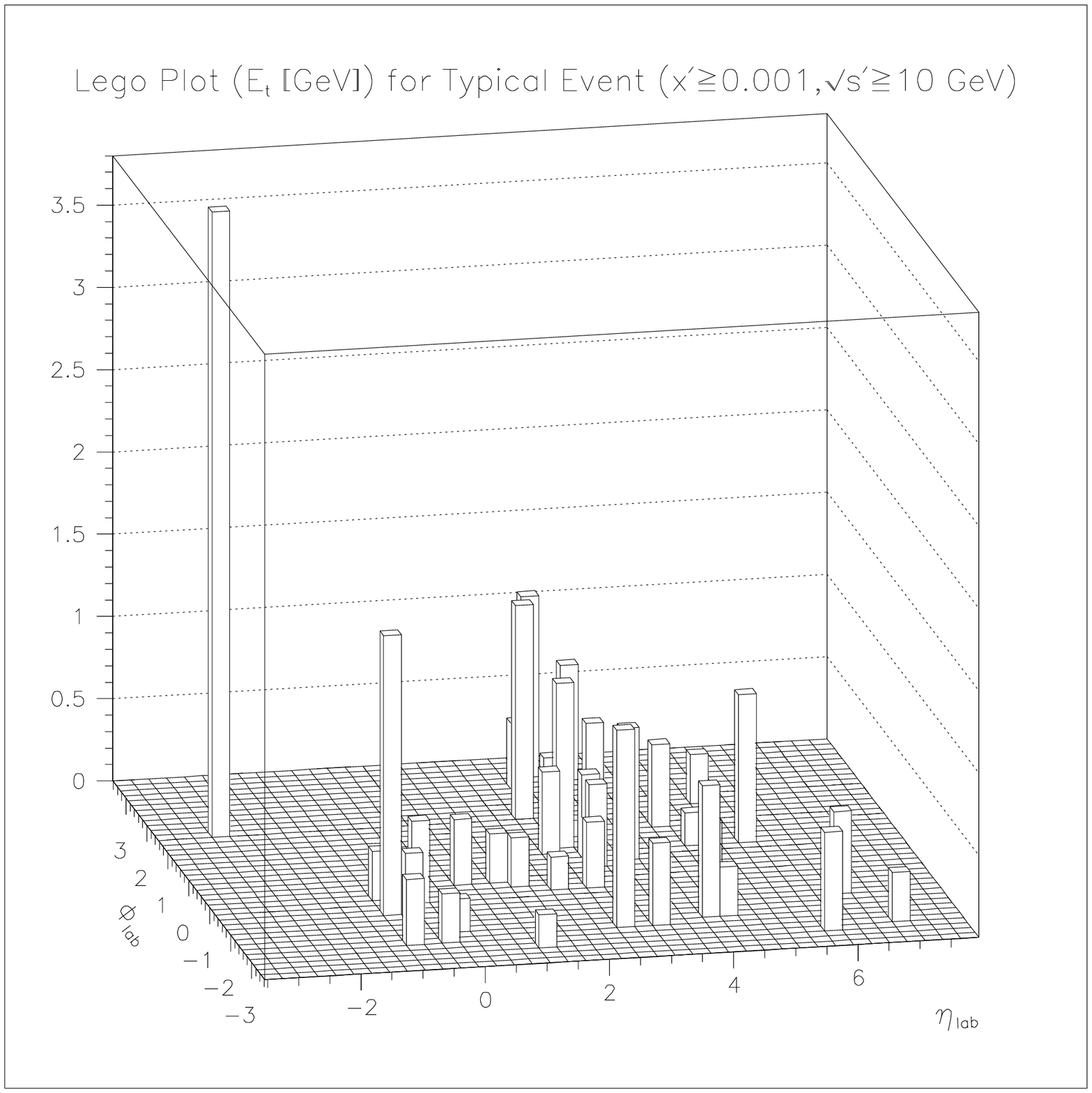}{80mm}{\em `Lego plot' for simulated QCD
instanton-induced event.}{legofig}

A general difficulty with instanton-induced processes is that they
are normally suppressed by large inverse powers of $Q^2$.  The cross
section is of the form
\beq\label{instcs}
\sigma(x^\prime ,Q^{\prime 2}) \sim
\frac{1}{s^\prime} \exp\left\{ -\frac{4\pi}{\alpha_s(Q^\prime)}
S(x^\prime)\right\}
\eeq
where $s^\prime, x^\prime, Q^\prime$ are the c.m.\ energy squared
and Bjorken variables of the instanton subprocess (the blob I in
\fref{instantfig}).  $S(x^\prime)$ is an increasing function,
approaching unity at large $x^\prime$ but unfortunately not yet
calculable at small $x^\prime$. This translates into a contribution
to the structure function $F_2$ as shown in \fref{lgf2p}.
The dashed curves are contours of constant $S$.
The region labelled ``data'' corresponds roughly to
the trend of the experimental data on $F_2$. Thus
at $x<0.3$ a detectable signal could be present, if
$S$ becomes small enough.

\ffig{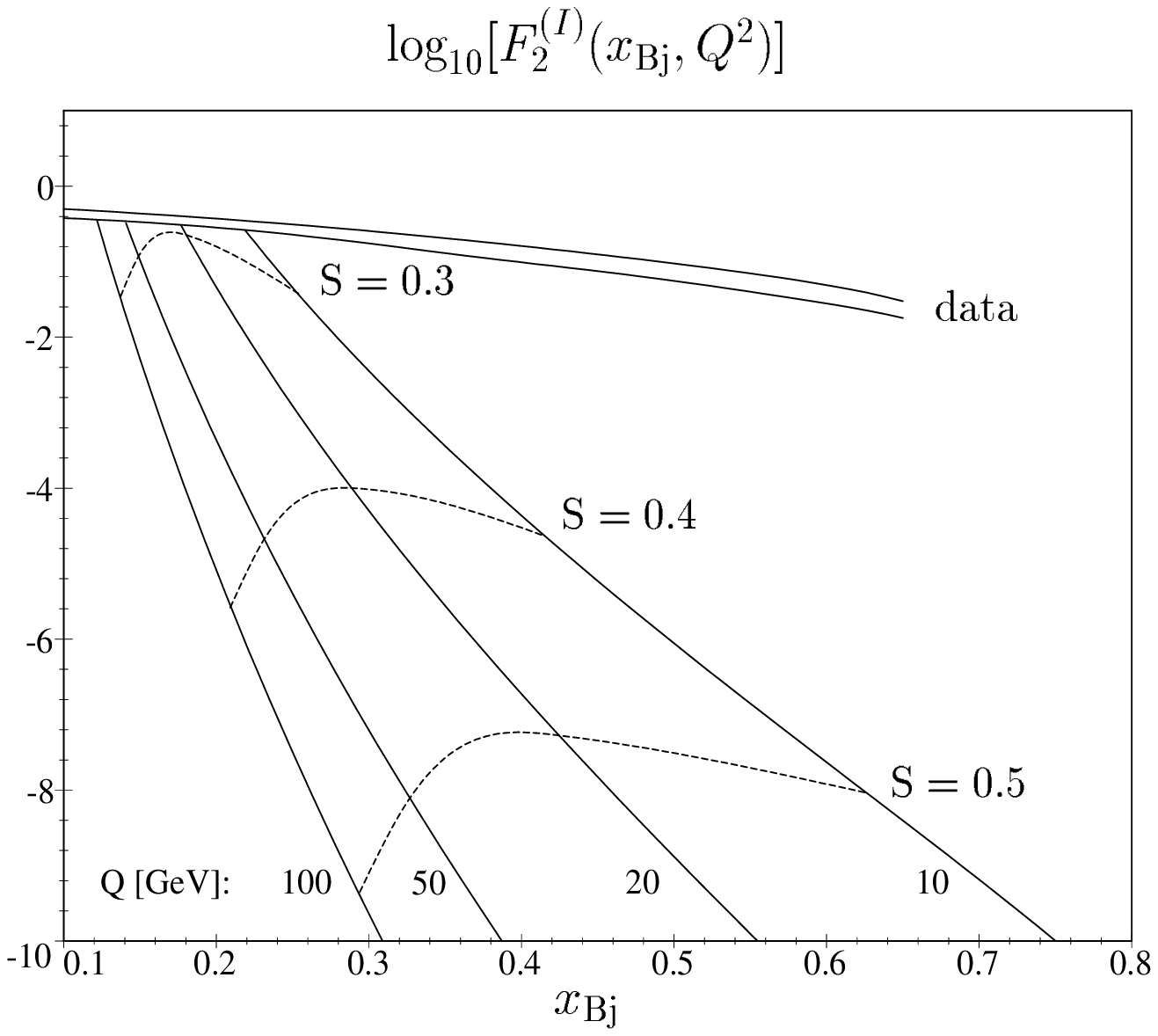}{80mm}{\em Logarithm of the instanton-induced
contribution to the structure function $F_2$ of the proton.}
{lgf2p}

\section{Conclusions}
Both experimental and theoretical studies of DIS final
states are still in their infancy.  Although things are not
so simple as in $\ee$ annihilation, because of the presence
of the incoming proton and its outgoing remnant, DIS has the
advantage of covering a wide range of $Q^2$ at a single beam
energy.  This is already leading to evidence for the running
of $\as$ from multi-jet production rates; we can expect similar
results from analyses of event shapes and jet fragmentation, plus
a host of new data on power-suppressed corrections, which will
improve understanding of the interface between perturbative and
non-perturbative QCD.  In the small-$x$ region, theoretical
ideas and experimental data on final state properties could be
crucial in elucidating the mechanism that causes the rise
in the structure function $F_2$.  A search for exotic final
states due to instanton-induced processes could shed new
light on non-perturbative QCD, and might establish the
credibility of analogous electroweak processes as a source
of baryon- and lepton-number violation within the Standard Model.
\begin{center}
{\large\bf Aknowledgement}
\end{center}
It is a pleasure to thank Violette Brisson and the committee for
organizing this most stimulating workshop.

\vspace{2cm}

\end{document}